# Non-planar geometrical effects on the magnetoelectrical signal in a three-dimensional nanomagnetic circuit

*Fanfan Meng[1], Claire Donnelly*[1], Claas Abert[2,3], Luka Skoric[1], Stuart Holmes[4], Zhuocong Xiao[5], Jung-Wei Liao[1], Peter J. Newton[1], Crispin H.W. Barnes[1], Dédalo Sanz-Hernández[1,6], Aurelio Hierro-Rodriguez[7,8], Dieter Suess[2,3], Russell P. Cowburn[1], and Amalio Fernández-Pacheco**[1,8]*

[1]Cavendish Laboratory, University of Cambridge, Cambridge UK
[2]Faculty of Physics, University of Vienna, Vienna, Austria
[3]Research Platform MMM Mathematics-Magnetism-Materials, University of Vienna, Austria
[4]London Centre for Nanotechnology, UCL, London, UK
[5]Nanoscience Centre, University of Cambridge, Cambridge, UK
[6]Unité Mixte de Physique, CNRS, Thales, Université Paris-Saclay, Palaiseau, France
[7]Depto. Física, Universidad de Oviedo, 33007 Oviedo, Spain
[8]SUPA, School of Physics and Astronomy, University of Glasgow, UK

* cd691@cam.ac.uk
** amalio.fernandez-pacheco@glasgow.ac.uk

**Expanding nanomagnetism and spintronics into three dimensions (3D) offers great opportunities for both fundamental and technological studies. However, probing the influence of complex 3D geometries on magnetoelectrical phenomena poses important experimental and theoretical challenges. In this work, we investigate the magnetoelectrical signals of a ferromagnetic 3D nanodevice integrated into a microelectronic circuit using direct-write nanofabrication. Due to the 3D vectorial nature of both electrical current and magnetisation, a complex superposition of several magnetoelectrical effects takes place. By performing electrical measurements under the application of 3D magnetic fields, in combination with macrospin simulations and finite**

**element modelling, we disentangle the superimposed effects, finding how a 3D geometry leads to unusual angular dependences of well-known magnetotransport effects such as the anomalous Hall effect. Crucially, our analysis also reveals a strong role of the non-collinear demagnetising fields intrinsic to 3D nanostructures, which results in an angular dependent magnon magnetoresistance contributing strongly to the total magnetoelectrical signal. These findings are key to the understanding of 3D spintronic systems and underpin further fundamental and device-based studies.**

Since the discovery of the anisotropic magnetoresistance (AMR) by Lord Kelvin in 1857[1], the fundamental investigation and exploitation of phenomena concerning the interplay between magnetism and electrical transport has seen incredible progress[2]. Indeed, pioneering studies of intrinsic effects originating from spin-orbit coupling in ferromagnetic materials[3] such as AMR and anomalous Hall effect[4] (AHE) have been followed by discoveries of the giant magnetoresistance[5] (GMR) and tunnel magnetoresistance[6] (TMR). These effects have underpinned the magnetic data storage revolution of recent decades[2]. Building upon this success, the field of spintronics in recent years has focused on control of spin states via electrical currents through the spin-transfer torque[7] (STT) effect, which has led to the recent development of non-volatile random-access memory (MRAM) devices[8]. All these advances, together with its role in today's digital world, make spintronics one of the most successful areas of nanotechnology[2]. Today, alternative forms of controlling the magnetic state via different mechanisms, *e.g.*, spin-orbit torques[9] (SOT), electric fields[10] and optical probes[11] are garnering much interest, with the prospect that future spintronic devices will impact a significant number of technological areas[8], including the emerging field of neuromorphic computing[12].

To meet the ever-increasing demands for new functionalities and more energy efficient devices, fundamental paradigm shifts are required. One of the most promising innovations involves the expansion of spintronics into three dimensions (3D)[13] which, with advantages such as higher density and enhanced device connectivity, offers a wealth of opportunities for 3D spintronics devices. With proposals ranging from the 3D magnetic racetrack memory[14] and the magnetic rachet[15] that represent alternative routes to ultra-high density, high-performance logic and memory devices, to 3D interconnected memristors for neuromorphic computing[12], 3D spintronics offers a highly efficient answer to the demands the field is currently facing. Experimentally, there has been a recent surge in progress including the observation of fast domain wall velocities in cylindrical nanowires[16], the demonstration of field-mediated controllable domain wall movement in 3D conduits[17] and extensive degeneracy[18] in frustrated nanowire lattices[19]. The higher degrees of freedom and surface-to-volume ratio of 3D nanostructures also make them desirable for sensing applications[13]. In this realm, there have already been first demonstrations, including magnetic imaging with high-aspect-ratio nanowires used as high-resolution MFM tips[20] and flexible position sensors making use of 3D nanomembranes[21].

As well as offering exciting prospects for devices, the introduction of 3D geometrical effects (*e.g.,* curvature and chirality) also provide opportunities for new physics[13,22]. These include predictions for new magnetic textures[23,24] and curvature-induced effects[25–28], as well as exotic dynamic behavior[16,29,30]. Although the experimental realization of such effects is challenging, recent developments in synthesis[19,31–34] and characterization techniques[24,35–40] have led to a number of stimulating first confirmations and discoveries of the potential of 3D nanomagnetism[13,22,41]. For example, non-reciprocal spin-wave propagation[29,42] has been observed in rolled-up nanomembranes, while magnetic chiral spin textures have been realized in double helices[43]. When combined, the fields of 3D spintronics and nanomagnetism offer great potential for new functionalities.

Before it becomes possible to fully exploit the potential of 3D spintronics, however, a fundamental understanding of the influence of the 3D geometry on the magnetotransport properties is needed. In this paper, we demonstrate the direct integration of a complex 3D magnetic nanostructure into a microelectronic circuit via direct-write nanoprinting and characterize the behavior of intrinsic magnetotransport effects such as the AMR and AHE in a 3D nanocircuit under the application of external magnetic fields. The efficient integration of the 3D magnetic structure, together with the vectorial nature of both current and magnetisation in 3D nanostructures, results in an unconventional superposition of different magnetotransport effects that are measured simultaneously. We separate these different contributions by taking advantage of symmetry arguments and their distinct angular dependence in response to magnetic fields. This allows us to understand the unexpected angular dependence of AHE due to the 3D geometry, and crucially reveal the strong influence of the non-colinear demagnetising field on the 3D magnetotransport via the magnon magnetoresistance (MMR). These fundamental insights are key for the future study of new spintronic effects in 3D magnetic nanostructures, as well as the realization of 3D spintronic technologies.

**Fabrication of 3D nanomagnetic circuits**

One of the key building blocks in 3D spintronics is the magnetic nanobridge. This device not only interconnects electrical and magnetic parts in the nanomagnetic circuit[44] but can also host magnetic domain walls (DWs)[45] and magnetic spin waves[46], to serve as both a memory element[14] and a logic gate[47], offering the possibility of new computing architectures in which the traditional boundaries between interconnects, memory and logic are eliminated[13]. A rendering of the bridge design investigated in this study is shown in **Figure 1a** where, in additional to the main conduction channel, two side-legs are introduced to allow standard four-probe measurements. These side legs are placed diagonally across the main channel so that both longitudinal and transverse magnetoelectrical signals can be measured simultaneously, therefore providing complementary information about the magnetic state of the device (as discussed later in the Magnetotransport measurements section). This arrangement promotes an efficient use of space on the substrate as only four planar pads are required to fully probe a high aspect ratio 3D circuit. It also improves the mechanical stability[48] of the device, by having rotational-symmetric leads at both sides of the main bridge. The enlarged bases of the bridge used here are designed to improve the electrical contact with the pads.

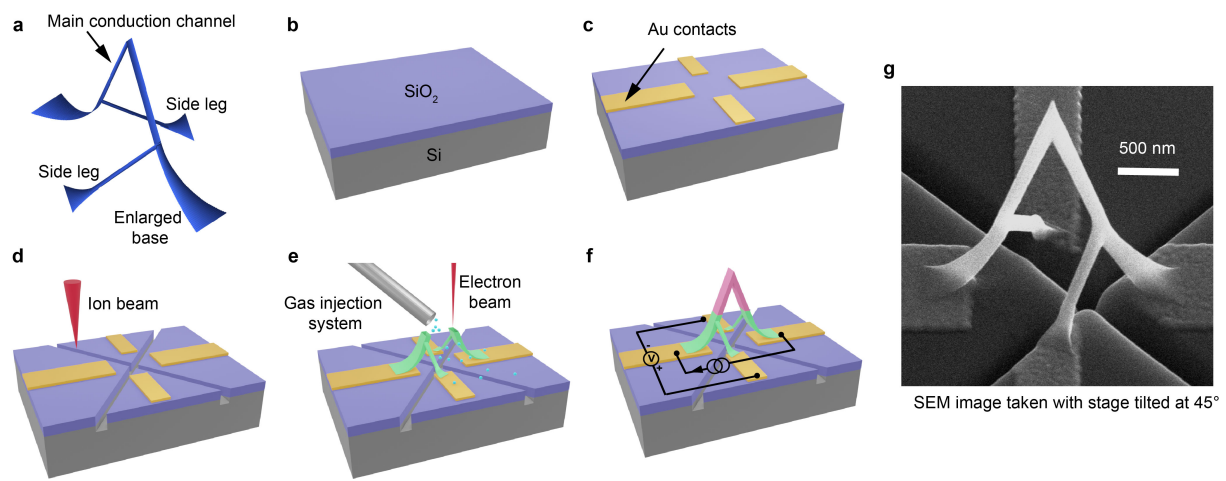

**Figure 1 | Integration of a ferromagnetic 3D nanobridge in a microelectronic circuit. a,** A rendering of the CAD design of the nanobridge investigated in the experiment. **b-e,** Steps followed to fabricate the nanobridge: **b,** preparation of a clean silicon dioxide substrate. **c,** patterning of electrical contacts by electron-beam lithography and electron-beam evaporation. **d,** milling of trenches by focused ion beam. **e,** 3D-nanoprinting of the bridge via focused electron beam induced deposition. **f,** Four-probe magneto-transport measurement configuration, where the voltage across the pink region is measured. **g,** SEM image of the as fabricated nanomagnetic circuit. Scale bar 500 nm. Image tilt 45°.

The realization of a 3D nanomagnetic circuit faces two challenges. The first challenge consists of creating an arbitrary 3D geometry with the desired material properties. The second challenge involves the integration of 3D magnetic structures into 2D microelectronics circuits. To overcome these issues, we employ focused electron beam induced deposition (FEBID), an additive fabrication technique with tens of nanometres in resolution[48–50]. Inspired by conventional 3D printers, recent developments in FEBID now make it possible to design beam scanning instructions of almost arbitrary 3D nanostructure geometries with varying curvatures and topologies, directly from standard 3D computer aided design (CAD) files[51]. In this way, with the appropriate use of precursor gases, 3D structures of high-quality ferromagnetic materials[52,53] can be fabricated directly on almost any substrate[48]. These capabilities make FEBID an ideal technique for the integration of a 3D magnetic circuit onto pre-patterned electrical contacts.

To begin the fabrication process of the 3D nanomagnetic circuit, four 50 nm thick gold contacts were patterned and deposited on a silicon substrate with a 300 nm thick silicon dioxide layer via electron beam lithography and electron beam evaporation (**Figure 1b-c**). Prior to FEBID 3D printing, trenches were milled by $Xe^+$ focused ion beam (FIB) between contacts to minimize the influence of conducting parasitic deposits[48], which could interfere during the transport measurements of the bridge (**Figure 1d**). Next, the nanobridge was directly printed on the four contacts via FEBID (**Figure 1e**) using dicobalt octacarbonyl [$Co_2(CO)_8$] as a precursor, under conditions which have been shown to result in greater than 95 at. % cobalt[52,53]. During the measurement, a constant current is supplied through the main leg of the bridge while the voltage is measured across the side leg contacts, as shown schematically in **Figure 1f**. A scanning electron microscopy (SEM) image of the resulting nanomagnetic circuit is shown in **Figure 1g**, demonstrating the successful connection of this complex, high aspect ratio 3D nanostructure to a planar circuit patterned on a substrate using well-defined leads. Detailed dimensions of the printed bridge and further information about the device are given in the Supporting Information.

**Magnetotransport measurements**

Following the realization of the 3D nanomagnetic circuit, we next consider magnetotransport (MT) measurements of the 3D nanostructure. The measurement setup used (**Figure 1f**) provides access to magnetoelectrical signals with different symmetries and, due to the 3D profile of the current (**Figure 2a**), a superposition of different MT effects was measured. To understand the contribution of different MT effects to the total signal probed in the device, and the influence of the 3D geometry, we performed measurements with the sample at different orientations with respect to the applied magnetic field direction. In this way, we can exploit the angular dependence and symmetries to separate the different MT effects[54].

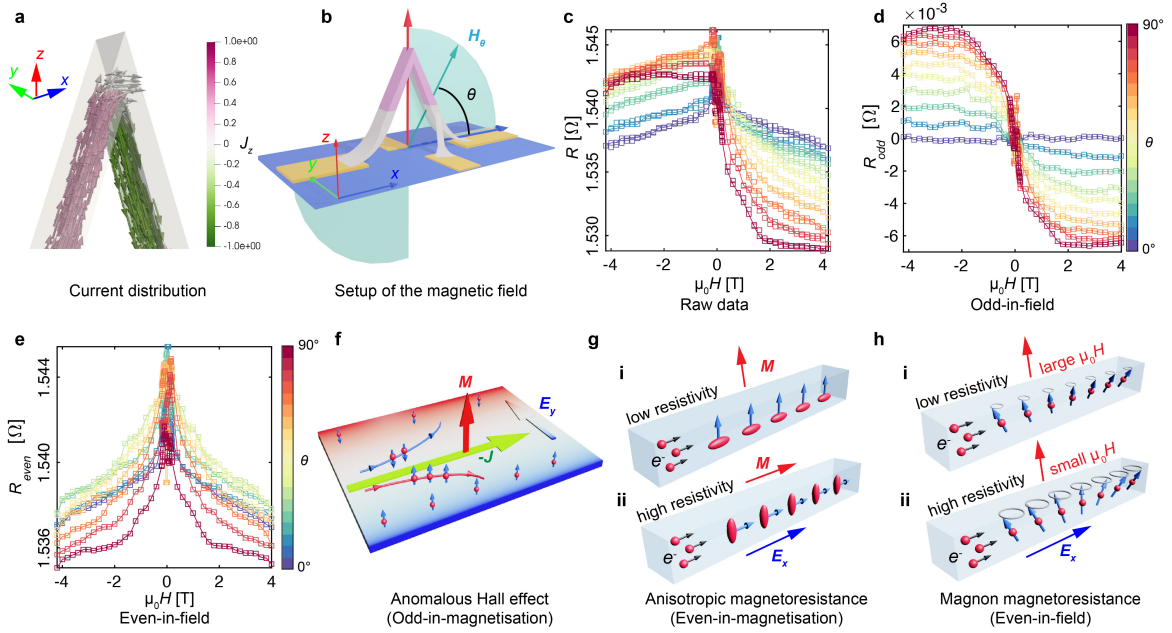

**Figure 2 | Magneto-transport measurements. a**, The simulation of the current density in the probed region of the 3D nanobridge with the colour indicating the *z* component of the current density. **b**, The schematic shows how the field is applied relative to the 3D nanobridge, $\theta$ is the angle between the applied field and the substrate. **c**, MT hysteresis loops obtained from -4 T to 4 T and 4 T to -4 T for each field angle $\theta$ from 0° to 90° (from violet to red). **d**, The odd-in-field signal, $R_{odd}$. **e**, The even-in-field signal, $R_{even}$. **f-h**, Schematics showing the main magnetotransport effects considered in the model: **f**, anomalous Hall effect: electrons with opposite spins are deflected in different directions due to the spin-orbit coupling mediated intrinsic and extrinsic (skew-scattering and side-jump) scattering-related mechanisms[4]. **g**, anisotropic magnetoresistance: variation in resistance induced by different degrees of scattering of spin-orbit coupled carriers[55]. **h**, magnon magnetoresistance: reduced resistance due to the suppression of spin waves by an applied magnetic field[56].

To probe the 3D response of the transport properties, we measure a full MT hysteresis loop from -4 T to 4 T and from 4 T to -4 T at 180 K, for fields applied between $\theta = 0°$ and $\theta = 90°$, with an interval of 10° in the *XZ* plane (**Figure 2b**), plotted from violet to red in **Figure 2c**. The raw data (**Figure 2c**) shows a clear angular dependence, with the signal becoming less symmetric as the angle changes from $\theta = 0°$ to $\theta = 90°$. In 2D, Hall bars are often patterned to separate various MT effects. Here, we instead make use of the different symmetries of the signal with respect to the sense of the applied field, to separate coexisting MT effects. Specifically, the raw data is first separated into the odd-in-field part as $R_{odd} = [R(\text{H}) - R(-\text{H})]/2$, and the even-in-field part as $R_{even} = [R(\text{H}) + R(-\text{H})]/2$ as shown in **Figure 2d** and **Figure 2e**, respectively. In this study, we focused on the high field range only, where the magnetisation is fully reversible, so that odd- and even-in-field signals correspond to odd- and even-in-magnetisation effects.

With the raw data separated into odd and even parts, we compare them to the symmetries and angular dependences of the anisotropic magnetoresistance (AMR), planar Hall effect (PHE), anomalous Hall effect (AHE) and ordinary Hall effect (OHE) on the current, internal magnetisation and magnetic field induction[3]:

$$\boldsymbol{E} = \rho_\perp \boldsymbol{J} + \underbrace{\overbrace{(\rho_\parallel - \rho_\perp)}^{\rho_G}[\boldsymbol{m}\cdot\boldsymbol{J}]\boldsymbol{m}}_{\text{AMR +PHE}} + \underbrace{\rho_{AHE}\boldsymbol{m}\times\boldsymbol{J}}_{\text{AHE}} + \underbrace{R_{OHE}\boldsymbol{B}\times\boldsymbol{J}}_{\text{OHE}} \quad (1)$$

where $\boldsymbol{E}$ is the electric field, $\boldsymbol{J}$ is the current density vector, $\boldsymbol{m}$ is a unit vector in the magnetisation direction, $\rho_\parallel$ is the resistivity for $\boldsymbol{J}$ parallel to $\boldsymbol{m}$, $\rho_\perp$ is the resistivity for $\boldsymbol{J}$ perpendicular to $\boldsymbol{m}$, $\rho_{AHE}$ is the anomalous Hall resistivity, $R_{OHE}$ is the ordinary Hall coefficient ($R_{OHE} = \rho_{OHE}/B$, where $\rho_{OHE}$ is the ordinary Hall resistivity, which is a function of $B$) and $\boldsymbol{B}$ is the total magnetic field induction, $\boldsymbol{B} = \mu_0(\boldsymbol{H}_a + \boldsymbol{H}_d + \boldsymbol{M})$. Here, $\boldsymbol{H}_a$ is the applied field, $\boldsymbol{H}_d$ is the demagnetising field and $\boldsymbol{M}$ is the magnetisation.

As the AHE is an odd-in-magnetisation effect (**Figure 2f**), the induced transverse electric field changes in sign with the reversal of magnetisation, and its strength depends on the component of magnetisation perpendicular to the current ($m_\perp$)[3]. The ordinary Hall effect is also an odd-in-field effect, and is usually a much smaller effect compared to the AHE[3]. The odd signal is plotted in **Figure 2d**, where first, we observe that for all $\theta$ values, $R_{odd}$ appears to level off for applied fields greater than 2 T. As the AHE dominates the odd signal and depends on the magnetisation only, this indicates that the magnetisation is effectively saturated at a field around 2 T. Above 2 T, a small negative slope can be observed (most significant at $\theta = 90°$), which we attribute to the ordinary Hall effect[52] (see below for more details).

We next consider the even-in-magnetisation effects - the AMR (**Figure 2g**) and PHE, which are the longitudinal and transverse components of the anisotropic resistivity and remain the same when the magnetisation reverses, with their magnitude depending on the magnetisation parallel to the current direction ($m_\parallel$)[3,54,55], as given by **Equation 1**. The even signal measured in the bridge is plotted in **Figure 2e**, where we notice that for all $\theta$, the resistance is always the highest when the applied field is around 0 T. This can be understood as the magnetisation at remanence tends to align along the long (easy) axes of the bridge due to shape anisotropy, which coincides with the current direction for this geometry. Since cobalt has a positive AMR ratio[52], the resistance is highest when the magnetisation and current directions are aligned. Secondly, in contrast to the odd signal, the even signal does not saturate at fields above 2 T but instead decreases further with applied field. As $(\rho_\parallel - \rho_\perp)$, *i.e.*, the AMR term in **Equation 1**, is not expected to change significantly after saturation[57], we attribute the measured change to the magnon magnetoresistance (MMR)[56,58], which has been reported as a linear and non-saturating negative MR present after all magnetic moments are fully saturated. As schematically shown in **Figure 2h**, this contribution is due to the progressive suppression of spin disorder caused by spin waves in a ferromagnet under an increasing field strength, which results in a drop of resistance due to a reduction in the electron-magnon scattering[56,58,59]. The magnitude of MMR depends on the strength, and not the sign of the applied field, and has therefore an even response with the applied field.

**Magnetotransport effects at high fields**

To obtain a quantitative understanding of the different contributions of the mentioned effects, we study the angular dependence of $R_{odd}$ and $R_{even}$ at high fields (± 4 T). We focus on the magnetoelectrical signals at high fields, where the magnetic state is close to uniform, and where the angular dependence of MT signals is usually the fingerprint of the underlying physical mechanisms[2]. In order to understand our measurements, we take into account the 3D nature of both the magnetisation and the current distribution, by making use of both a multi-macrospin model and a finite element method (FEM) analysis, as explained below.

First, we determine the magnetic state of the nanobridge at ± 4 T for different magnetic field directions, by modelling the magnetic configuration using an adapted multi-macrospin (multiple single domain) approximation: the probed region of the structure is considered to be made up of three single-domain sections, as marked in green, pink and yellow, respectively in **Figure 3a**. The interaction between the three regions is not considered in the model, *i.e.*, the magnetisation vectors for each section, $\boldsymbol{m_1(\theta)}$, $\boldsymbol{m_2(\theta)}$ and $\boldsymbol{m_3(\theta)}$ are determined independently, by minimizing the Zeeman and magnetostatic energies of each section at a given $\boldsymbol{\theta}$, where the shape anisotropy of each part is included as an independent demagnetising term. Due to the nanocrystalline nature of FEBID Co under these growth conditions[52], which results in magnetic properties dominated by shape anisotropy, we do not consider the intrinsic magnetocrystalline anisotropy of cobalt in the model. As described below, this approach is sufficient to fully understand the magnetic behavior of the nanocircuit from MT signals under the application of high magnetic fields. The detailed calculation is described in the Supporting Information.

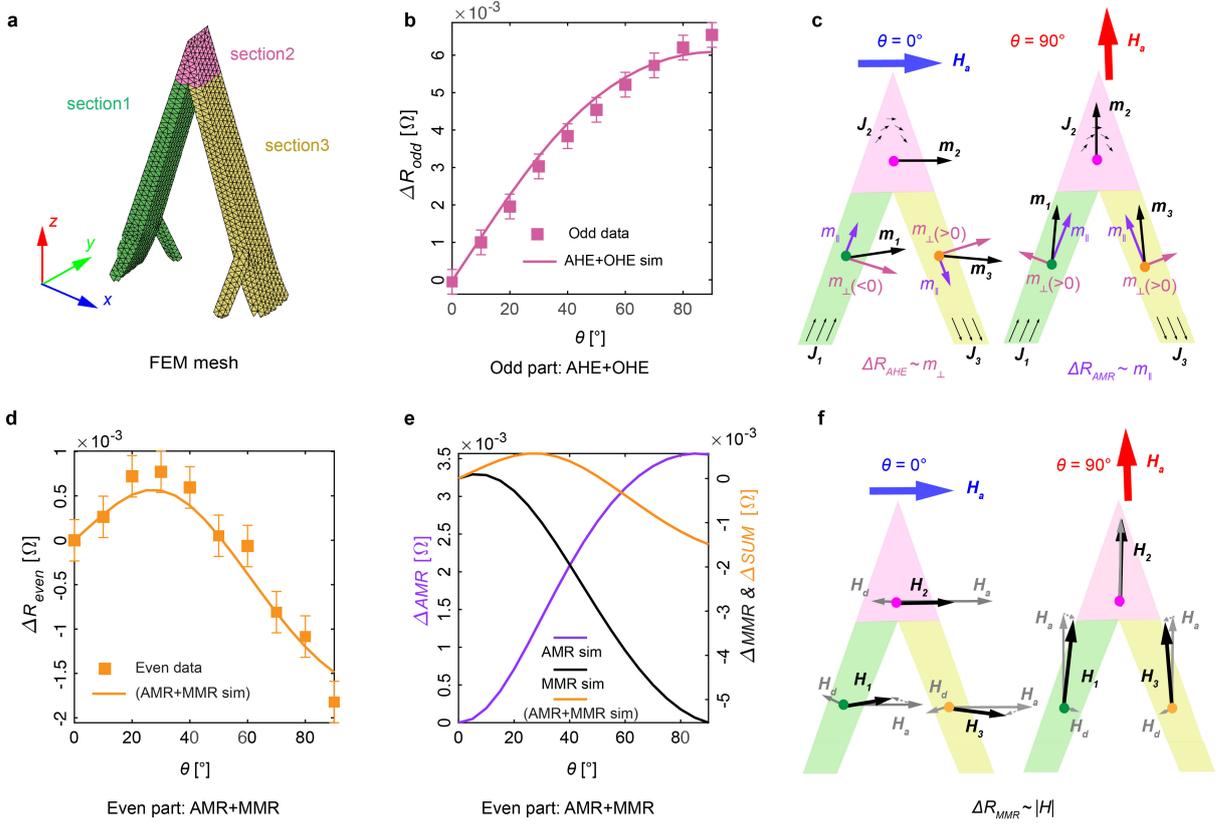

**Figure 3 | Analysis of resistance data at 4 T. a,** The FEM mesh of the bridge is divided into three domains. **b,** Comparison between the angular dependence of the odd part data and the AHE and OHE simulation. **c,** Averaged magnetisation vectors, $m_1(\theta)$, $m_2(\theta)$ and $m_3(\theta)$ for the three sections for $\theta = 0°$ and $\theta = 90°$ and their components on the current and current normal directions. **d,** Comparison between the angular dependence of the even part of the data and the sum of AMR and MMR simulation. **e,** Simulation of the AMR, MMR effects and their sum. **f,** Applied field $H_a$, demagnetising field $H_d$ and their vector sum $H$ for each section of the bridge.

After obtaining the magnetic configuration from the macrospin model, we simulate the MT signal by solving the electric potential $u$ across the side contacts using a finite element method (FEM) with a CAD-based FEM mesh that reproduces the dimensions of the printed nanobridge (**Figure 3a**). The influence of different MT effects is summarized as a magnetisation- and field-dependent resistivity tensor $\rho(m, B)$ which can be obtained from **Equation 1** (see Methods section). For each section and angle $\theta$ considered, a unique resistivity tensor is calculated from the modelled magnetisation distribution $m(\theta)$ and total field distribution $B(\theta)$, and assigned to the corresponding sections of the nanobridge. We note that a non-magnetic conducting layer underneath the bridge due to known Co parasitic deposition of cobalt[60] has been included in the simulations for a better quantitative agreement of the base resistance. Details of the FEM simulation setup are given in the Supporting Information. After obtaining the potential difference between side contacts for fields applied in all directions, the odd part and even part of the simulated results can be separated in the same way as for the experimental data, for comparison.

We first consider the angular dependence of the odd signal by plotting the average of $|R_{odd}(4\text{ T})|$ and $|R_{odd}(-4\text{ T})|$ with respect to the $\theta = 0°$ case (squares in **Figure 3b**), along with the simulated odd signal (line in **Figure 3b**), which is dominated by the AHE (see Supporting Information). We observe a continuous increase in the magnitude of the odd component as the field rotates from $\theta = 0°$ to $\theta = 90°$, and a good agreement between the simulations and the data, confirming that odd signal is due to combination of AHE and the smaller OHE[52]. We first consider the dominant contribution of AHE. The trend of an increasing AHE magnitude with increasing $\theta$ can be understood intuitively by considering the $\theta = 0°$ and $\theta = 90°$ extreme cases in **Figure 3c**, where the magnetisation vector *m* calculated for each section from the macrospin model is shown. The component of the magnetisation on the current direction ($m_\parallel$) and on the current-normal direction ($m_\perp$) are also plotted. As the AHE depends on $m_\perp$, at a first glance it might appear that the $\theta = 0°$ case would result in a larger AHE effect, as the magnitude of $m_\perp$ is larger for green and yellow sections. However, the geometry of the nanobridge results in opposite signs of $m_\perp$ for these two sections in the frame of reference of the current, and thus the two AHE signals cancel out. Using analogous arguments for the pink region, the current turns its direction along this section, also resulting in a negligible AHE signal. The opposite scenario occurs at $\theta = 90°$, where the same sign of $m_\perp$ for all three sections leads to the signals adding up. The 3D current distribution and contact arrangement thus explain why the AHE is greatest at $\theta = 90°$, and constitutes a demonstration of how vectorial current distributions flowing in 3D magnetic geometries may lead to deviations of the angular dependence of MT signals from familiar patterns observed in planar systems. The magnitude of OHE follows the same trend, which explains why the negative linear slope is the most obvious at $\theta = 90°$ in **Figure 2d**. From this analysis, we also obtain the AHE and OHE resistivities of the device. From the model fitting, we find $\rho_{AHE} = 5.6 \times 10^{-9}$ Ωm and $R_{OHE} = -1.2 \times 10^{-10}$ Ωm/T, of the same range as the anomalous Hall resistivity and ordinary Hall coefficient reported in the literature for FEBID-deposited cobalt[52].

We next consider the even part of the signal, with the average of $R_{even}(4\text{ T})$ and $R_{even}(-4\text{ T})$ with respect to the $\theta = 0°$ case, plotted as squares in **Figure 3d**. We observe a peak at around $\theta = 30°$, as well as an increased resistance at $\theta = 0°$ compared to $\theta = 90°$.

We first examine what are commonly considered the main intrinsic contributions to the even signal: the AMR and PHE. The simulated sum of AMR and PHE signals (referred for brevity as AMR) is plotted in **Figure 3e** (purple line), which takes the form of a monotonic increase in the resistance with increasing $\theta$. Again, this angular dependence can be understood by considering $m_\parallel$ at $\theta = 0°$ and $\theta = 90°$ in **Figure 3e**: $m_\parallel$ is larger at $\theta = 90°$, so we would expect the resistance to be larger at $\theta = 90°$ due to AMR and PHE, consistent with the simulated signal. However, the experimental data in **Figure 3d** exhibit a very different angular dependence, implying that the even data cannot be fully explained by these two effects.

To understand the significant difference in the even part of the signal, we consider the angular dependence of magnon magnetoresistance as an additional contribution. The MMR results in a change of resistivity that can be described by the electron-magnon scattering model developed by Raquet et al[56,58,61]:

$$\Delta\rho_{mmr}(T,B) \approx \rho(T,B) - \rho(T,0) \propto \frac{BT}{D(T)^2} \ln\left(\frac{\mu_{Bohr}B}{k_BT}\right) \qquad (2)$$

where $T$ is the temperature, $D(T)$ is the temperature dependent magnon stiffness, $\mu_{Bohr}$ is the Bohr magneton, $k_B$ is the Boltzmann constant, and $B$ is the projection of the total effective magnetic field, $\boldsymbol{B} = \mu_0(\boldsymbol{H} + \boldsymbol{M})$, on the direction of the magnetisation, *i.e.*, the magnitude of the total effective field, acting to suppress the magnitude of spin-waves present in the system. Here $\boldsymbol{H}$ is the vector sum of the applied field $\boldsymbol{H_a}$ and demagnetising field $\boldsymbol{H_a}$, $\boldsymbol{H} = \boldsymbol{H_a} + \boldsymbol{H_d}$, and $\boldsymbol{M}$ is the magnetisation.

For a constant temperature, **Equation (2)** leads to a negative change of resistivity that decreases almost linearly with the magnitude of the effective field, which is consistent with the even part of the experimental signal for applied fields greater than 2 T (**Figure 2e**). Although $\Delta\rho_{mmr}$ is not dependent on the direction of the magnetisation with respect to the current[56,58], this does not necessarily mean that no change in the $\Delta\rho_{mmr}$ will result from different directions of the applied magnetic field. Previous studies investigating the contribution of MMR in nanostructures have mainly focused on measurements under fields applied along the easy axis of 2D thin films or nanowires[56,58], where the demagnetising field is negligible, leading to an effective field equivalent to the applied field[56,58]. However, in the case of a 3D nanocircuit such as the one studied here, the non-planar geometry results in an applied field always oblique to at least one section of the circuit. This results in a non-zero demagnetising field modifying the effective magnetic field at any angle.

In order to compare the experimental MMR with simulations, we calculate the nonzero demagnetising field as a function of $\theta$ from the macrospin model, as described in the Supporting Information. The resulting change of resistivity, $\Delta\rho_{mmr}$, plotted as a black line in **Figure 3e**, is obtained from **Equation (2)**. An opposite trend in the angular dependence of MMR with respect to the AMR is observed, with the resistance becoming more negative with increasing angle. Again, we consider the extreme cases of $\theta = 0°$ and $\theta = 90°$ in **Figure 3f** to understand this angular dependence intuitively. At $\theta = 0°$, the applied field has a larger component perpendicular to each section's easy axis, resulting in a larger demagnetising field, and thus a lower magnitude of the overall effective magnetic field, leading to a higher resistance. At $\theta = 90°$, the field is more aligned with the easy axes for all three sections, leading to a lower demagnetising field, and a larger magnitude of the total effective field, associated with a larger drop of resistivity, as seen in the simulated data.

Finally, we compare the sum of AMR, PHE and MMR obtained from simulations (**Figure 3d-e**, orange line), to the even data (**Figure 3d**, squares). An excellent agreement with experiments is observed, with the maximum resistance at around $\theta = 30°$, and the overall angular trend well reproduced. This demonstration of the strong influence of the three-dimensional geometry on the magnetotransport reveals the importance of non-collinear alignments between magnetic fields and geometry in non-planar magnetic nanocircuits. In particular, this work demonstrates how magnetostatic interactions in 3D geometries manifest through a significant deviation of the MMR contribution.

**Conclusions**

We have investigated the magnetoelectrical response of a 3D nanomagnetic circuit created by advanced 3D nano-printing. By exploiting signal symmetries with respect to magnetic field configurations, we were able to address the superposition of different magnetotransport effects. Specifically, we combined electrical measurements with finite-element calculations to disentangle and understand key magnetotransport effects (Hall effect and magnetoresistance signals) within the nano-circuit, obtaining a clear understanding of their magnitudes and angular dependences as a function of external magnetic fields applied along multiple directions.

In this way, we identified that the 3D geometry of the magnetic nanostructure has a major effect, inducing deviations of the Hall effect signal from the angular dependence usually observed in planar geometries. The 3D vector nature of both *m* and *J* is responsible of this unusual angle dependence, due to the fact that signals that, *e.g.*, cancel out in planar geometries may add up in 3D. Moreover, the overall magnetoelectrical signal has a significant angular-dependent magnon contribution, due to varying magnetostatic interactions throughout a 3D circuit which are not present in a standard planar magnetic device.

The new insights into the influence of a 3D geometry on the magnetotransport effects reported here provide the basis for exploring new spintronic effects emerging in three dimensions[27,28,62] and long-term, the realization of 3D devices. The methodology shown here combining FEBID 3D printing with standard planar lithography can be extended to more complex 3D geometries and other materials, opening the door to the fundamental study of new phenomena that exploit the interplay between 3D geometry and magnetotransport. 3D spintronic effects may find key applications in the future of areas such as magnetic computing based on nanomagnetic logic[15,44], domain wall[14,47] and skyrmion devices[63], magnonics[25,46,64], magnetic neural networks[12] and frustrated magnetic systems such as artificial spin-ice systems[19,65,66].

**Methods**

**Fabrication** The electrical contacts on which the 3D nanobridge was printed were patterned by electron-beam lithography followed by electron-beam evaporation of 5 nm Cr/50 nm Au. The milling of the trenches was performed using a Xenon Plasma focused ion beam microscope using 74 pA current and 30 kV acceleration voltage. The 3D nanostructure was printed by the same microscope using 340 pA current, 30 kV acceleration voltage, and $CO_2(CO)_8$ as the precursor. The dwell points and dwell time were calculated using the algorithm developed by Skoric et al[51].

**Measurements** The magneto-transport measurements were performed in a bath flow Helium cryostat at a constant temperature of 180 K. The standard four-terminal AC lock-in technique employed a constant amplitude current input of 0.6 µA at a frequency of 33 Hz.

**FEM calculations of MR** The influence of different MT effects is summarized as a magnetisation-dependent resistivity tensor $\rho(m, B)$ which is reformulated from **Equation (1)** as

$$E = \rho(m, B)J$$

$$\rho(m, B) = \begin{pmatrix} \rho_\perp + \Delta\rho_{mmr} + \rho_G m_x^2 & \rho_G m_x m_y - \rho_{AHE} m_z - R_{OHE} B_z & \rho_G m_x m_z + \rho_{AHE} m_y + R_{OHE} B_y \\ \rho_G m_x m_y + \rho_{AHE} m_z + R_{OHE} B_z & \rho_\perp + \Delta\rho_{mmr} + \rho_G m_y^2 & \rho_G m_y m_z - \rho_{AHE} m_x - R_{OHE} B_x \\ \rho_G m_x m_z - \rho_{AHE} m_y - R_{OHE} B_y & \rho_G m_y m_z + \rho_{AHE} m_x + R_{OHE} B_x & \rho_\perp + \Delta\rho_{mmr} + \rho_G m_z^2 \end{pmatrix}$$

where $m_x$, $m_y$, and $m_z$ are the *x*, *y* and *z* component of the magnetisation vector **m**, $B_x$, $B_y$ and $B_z$ are the *x*, *y* and *z* component of the total magnetic induction **B** and $\Delta\rho_{mmr}$ is the change of resistivity due to the MMR effect.

The conductivity tensor is then obtained by inverting the resistivity tensor, $\boldsymbol{\sigma} = \boldsymbol{\rho}^{-1}$. For each section, and each field angle $\theta$, a unique conductivity tensor is calculated from the modelled magnetisation distribution **m** *(θ)* and assigned to the corresponding sections of the nanobridge. The electric potential *u* at the side contacts is then computed by solving the partial differential equation, $\nabla \cdot [-\sigma(\nabla u)] = 0$, using a finite element method implemented with a CAD-design-based finite element mesh that reproduces the dimensions of the printed nanobridge as shown in **Figure 3d**. More information on the setup of the FEM simulation is given in the supporting information.

**Acknowledgements**

This work was funded by EPSRC Early Career Fellowship EP/M008517/1 and the Winton Program for the Physics of Sustainability. F.M. was supported by China Scholarship Council. C.D. was supported by the Leverhulme Trust (ECF-2018-016), the Isaac Newton Trust (18-08), and the L'Oréal-UNESCO U.K. and Ireland Fellowship For Women In Science. L.S. was supported by the EPSRC Cambridge NanoDTC EP/L015978/1. A.H.-R. acknowledges the support from European Union's Horizon 2020 research and innovation program under Marie Skłodowska-Curie grant ref. H2020-MSCA-IF-2016-746958 and from Spanish AEI under project reference PID2019–104604RB/AEI/10.13039/501100011033. We are grateful to Oksana Chubykalo-Fesenko, Can Avci and Eduardo Martínez for fruitful discussions.

**Competing Interests statement**

The authors declare no competing interests.